\PassOptionsToPackage{polish,english}{babel}
\documentclass[a4paper,11pt]{article}
\usepackage{pos}
\usepackage{amsmath,amsfonts,verbatim,graphicx,float}
\usepackage{bm}
\usepackage{polski}
\usepackage[utf8]{inputenc}
\usepackage[polish, english]{babel}
\newcommand*\difff{\mathop{}\!\mathrm{d}}

\title{Simulations of Cosmic Ray Ensembles originated nearby the Sun}
 \ShortTitle{Simulations of CREs originated nearby the Sun}

\author*{David Alvarez-Castillo}
\author{Piotr Homola, Dariusz Gora, Dhital Niraj, \\ Gabriela Opi{\l}a, Justyna M\k{e}drala, Bo\.zena Poncyljusz,}

\forColl{CREDO}  

\emailAdd{dalvarez@ifj.edu.pl}

\abstract{Cosmic Ray Ensembles (CRE) are yet not observed groups of cosmic rays with a common primary interaction vertex or the same parent particle. One of the processes capable of initiating identifiable CRE is an interaction of an ultra-high energy (UHE) photon with the solar magnetic field which results in an electron pair production and the subsequent synchrotron radiation. The resultant electromagnetic cascade forms a very characteristic line-like front of a very small width ($\sim$ meters), stretching from tens of thousands to even many millions of kilometers. In this contribution we present the results of applying a toy model to simulate detections of such CRE at the ground level with an array of ideal detectors of different dimensions. The adopted approach allows us to assess the CRE detection feasibility for a specific configuration of a detector array. The process of initiation and propagation of an electromagnetic cascade originated from an UHE photon passing near the Sun, as well as the resultant particle distribution on ground, were simulated using the CORSIKA program with the PRESHOWER option, both modified accordingly. The studied scenario results in photons forming a cascade that extends even over tens of millions of kilometers when it arrives at the top of the Earth's atmosphere, and the photon energies span practically the whole cosmic ray energy spectrum. The topology of the signal consists of very extended CRE shapes, and the characteristic, very much elongated disk-shape of the particle distribution on ground illustrates the potential for identification of CRE of this type.}

\FullConference{37$^{\rm{th}}$ International Cosmic Ray Conference (ICRC 2021)\\
		July 12th -- 23rd, 2021\\
		Online -- Berlin, Germany}

\begin{document}
\maketitle

\section{Introduction}

Within this work we consider electromagnetic cascading of ultra-high-energy (UHE) photons transversing regions nearby the Sun. We present the results of simulations of this process, from the origin of the corresponding shower to the propagation down to the ground level on Earth as well as the simulated detected signal. The electromagnetic cascades front is expected to be an extended thin particle distribution that might extend from thousands to millions of kilometers. The simulations reveal that the topology of footprint on Earth consists of an extended line with a rather elliptical or circular core depending on the angle of incidence. In order to provide a good characterization of the distribution, special care of the appropriate physics of the cascade development and particle tracking is taken in the simulations. The cascading process consists of magnetic pair production in which the produced electron-positron pairs will emit photons through Bremsstrahlung. Moreover, those produced photons bearing high energies will in turn undergo magnetic pair-production and repeat the process. The resulting cascade nearby the Sun therefore comprises thousands of photons and several $e^{+}e^{-}$, whose extended distribution develops while propagating towards the Earth, traveling  a distance  of $\sim 1.5\times10^{11} \ \mathrm{m}$. These \textit{superpreshowers} (SPS) carry many more secondary particles that \textit{preshowers} originated in the geomagnetic field that comprise only a few hundred particles with a very narrow spatial distribution, of less than 1 m.

Our study is an alternative approach to the standard searches of UHE photons that might be produced from the decay of dark matter particles~\cite{PhysRevLett.107.131302, Ahmed:2009zw, PhysRevLett.112.091303} or produced by the Greisen-Zatsepin-Kuzmin (GZK) effect~\cite{PhysRevLett.16.748, Zatsepin:1966jv}, i.e., high energy particle or ions interacting with the cosmic microwave background, as seen from the steepening of the cosmic rays spectrum about the $ 4\times 10^{19}\mathrm{\ eV}$ energy value.

The main purpose of this work is to demonstrate the feasibility of detection of CREs on the surface of the Earth. We use the simulated preshower cascade as an input into the simulation for detectors at the ground level and choose a fixed array of detectors with variable size. Such variation allow us to set up threshold for identification of the CRE footprint under ideal conditions of total detector efficiency. The obtained results represent a step forward towards optimization of global CRE detection within the CREDO initiative~\cite{CREDO:2020pzy}.

\section{Simulations}

\begin{figure}%
    \centering
 \includegraphics[width=1.0\textwidth]{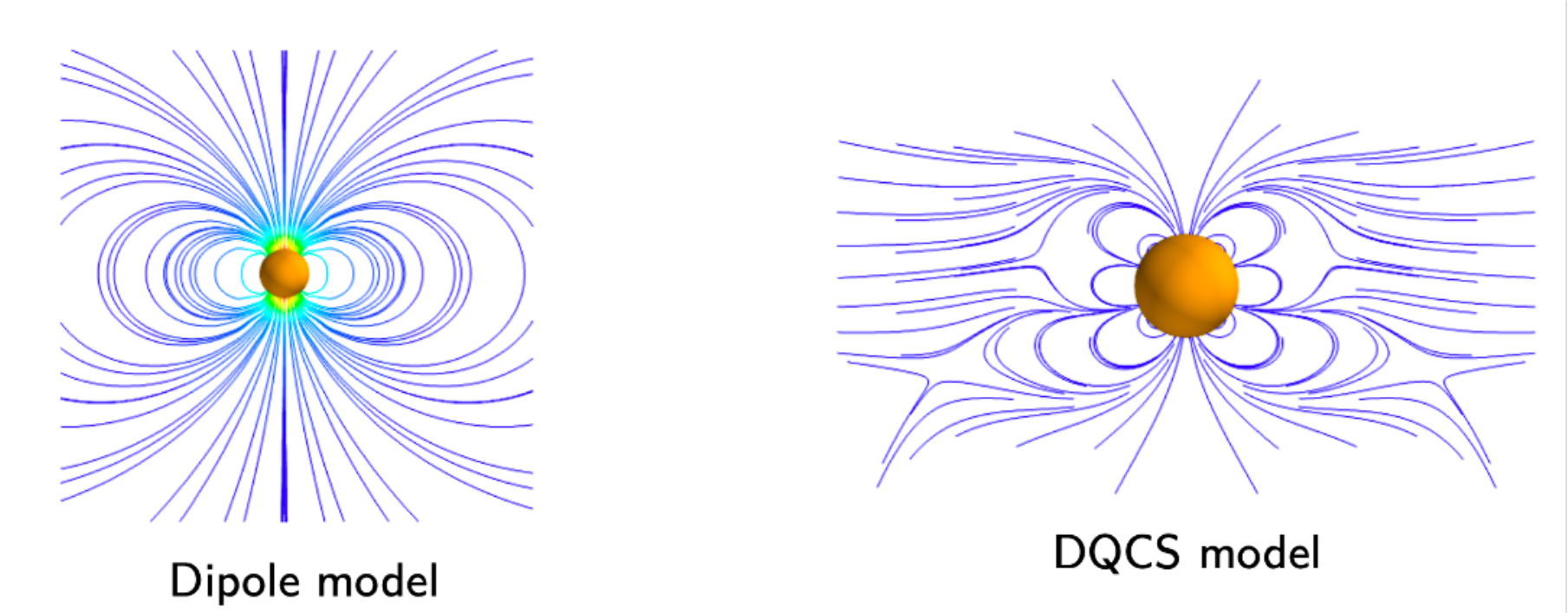}%
    \caption{Models of Sun's magnetic field: simple dipole and dipole-quadrupole-current-sheet (DQCS)~\cite{dqcs}.}%
    \label{fig:magnetic}%
\end{figure}

The simulations in this work were carried out using the CORSIKA program with the option of loading PRESHOWER. Both codes were modified in order to meet our chosen configurations and physical conditions.
When considering the contribution of the magnetic field of the Sun to the SPS  effect within our simulations, we choose either a simple dipole approximation or the so called dipole-quadrupole-current-sheet (DQCS) model~\cite{dqcs}, see figure~\ref{fig:magnetic} for a schematic representation. Within the dipole approach, the value of the magnetic moment that produces the magnetic field used in the simulation is $6.87 \times 10^{32} \mathrm{\ G \cdot cm^{3}}$~\cite{Bednarek:1999wg}. The simplicity of this model allows for variation of the orientation of the dipole in order to study its effect on the SPS distribution of particles that arrive at the top of the Earth's atmosphere. Alternatively, the DQCS model which is more realistic,  allows for a better description of the $e^+ e^-$ that will arrive on Earth, and an improved treatment of  the magnetic Bremsstrahlung process. Using the PRESHOWER code~\cite{2005CoPhC.173...71H} which includes all the necessary physical processes together with the pair production formalism of reference~\cite{RevModPhys.38.626} we compute the number of $e^{+} e^{-}$ pairs produced $n_\mathrm{pairs}$ from UHE photons $n_\mathrm{photons}$ interacting with the magnetic field $H$ as
\begin{equation}
n_\mathrm{pairs} = n_\mathrm{photons} \{1 - \exp\left[-\alpha \left(\chi\right) \difff l \right] \} \mathrm{,} \label{eqn:n_pairs}
\end{equation}
with $\difff l$ as the corresponding path length and a photon attenuation coefficient $\alpha\left(\chi\right)$ with $\chi \equiv \frac{1}{2} \frac{h \nu}{m_{\mathrm{e}} c^{2}} \frac{H}{H_\mathrm{cr}}$ and $H_\mathrm{cr} \equiv \frac {m_{\mathrm{e}}^{2} c^{3}}{e \hbar} = 4.414 \times 10^{13} \mathrm{\ G}$. In case that $H \ll H_\mathrm{cr}$, which corresponds to the ultra-relativistic limit, $\alpha\left(\chi\right)$ is given by
\begin{equation}
\alpha\left(\chi\right) = \frac{1}{2} \frac{\alpha_\mathrm{em}}{\lambdabar_\mathrm{c}} \frac{H}{H_\mathrm{cr}} T\left(\chi\right) \mathrm{,}
\end{equation}
with ${\lambdabar_\mathrm{c}}$ as the electron Compton wavelength, $T\left(\chi\right) \simeq \frac{0.16} {\chi} K_{1/3}^{2} \left(\frac{2}{3\chi}\right) \mathrm{,}$ and $K_{1/3}$  being a modified Bessel function. The probability of conversion of an UHE photon into a $e^{+} e^{-}$ pair within the interval $\difff l$  can be approximated as
\begin{equation}
p_\mathrm{conv} = 1 - \exp\left(-\alpha \left(\chi\right) \difff l\right) \simeq \alpha \left(\chi\right)\difff l \mathrm{,} \label{eqn:p_conv}
\end{equation}
or for a very large distance $L$,
\begin{equation}
P_\mathrm{conv} = 1 - \exp[ - \int\limits_{0}^{L} \alpha \left(\chi\right)\difff l] \mathrm{.} \label{eqn:P_conv}
\end{equation}
Each pair member carries an energy fraction which follows the distribution
\begin{equation}
\frac{\textrm{d}n}{\textrm{d}\varepsilon} \approx \frac{\alpha_\mathrm{em} H}{\lambdabar_\mathrm{c}} \frac{\sqrt{3}}{9\pi \chi} \frac{[2+\varepsilon(1-\varepsilon)]} {\varepsilon(1-\varepsilon)}
K_{\frac{2}{3}}\left[\frac{1}{3 \chi \varepsilon(1-\varepsilon)}\right],
\end{equation}as presented in~\cite{1983ApJ...273..761D}.
Time and space tracking of particles in the cascade is necessary for determination of arrival times and lateral distribution at the top of the Earth's atmosphere. Standard electrodynamics are obeyed for a particle with kinetic energy $E$, charge $q$, and direction  $\bm{\hat{v}}$ in a region of magnetic field $\bm{B}$,
\begin{eqnarray}
f(y) = \frac{9\sqrt{3}}{8 \pi} \frac{y}{( 1 + \xi y )^3}\left\{\int\limits_{y}^{\infty} K_{\frac{5}{3}}(z)\difff z 
+\frac{(\xi y)^{2}}{1 + \xi y}  K_\frac{2}{3} (y)\right\}\mathrm{\quad}
\end{eqnarray}
with $\xi = \frac{3}{2} \frac{H_\perp}{H_\mathrm{cr}} \frac{E}{m_\mathrm{e} c^2}$, $E$ and $m_\mathrm{e}$ as the electron energy and rest mass respectively and $y$ which is related to the 
emitted photon energy $h \nu$
\begin{equation}
y\left(h \nu\right) = \frac{ h \nu}{\xi \left( E - h \nu \right)} \mathrm{.}
\end{equation}
Furthermore, the associated Bremsstrahlung process probability is given by
\begin{equation}
P_\mathrm{brem}\left(B_\perp, E, h\nu, \difff l\right) = \difff l \int\limits_{0}^{E} I\left(B_\perp, E, h\nu\right) \frac{ \difff\left(h\nu\right)}{h\nu} \mathrm{,} 
\end{equation}
where
\begin{equation}
I\left(B_\perp, E, h\nu\right) \equiv \frac{h\nu \difff N}{\difff \left(h \nu\right) \difff l} \mathrm{,}
\end{equation}
with the number of photons $\difff N$ with energy between $h\nu$ and $h \nu + \difff \left(h \nu \right)$  which are emitted over $\difff l$. For more details on this computation and implementation in the simulation we refer the reader to~\cite{Dhital:2018auo}.

\section{Results}

For the simulations performed in this study we have implemented diverse conditions related to the variation of model parameters, like the impact parameter $R$ for the cascade production nearby the Sun, or the consideration of polar or equatorial incidence of particles. We find that the conversion probability for magnetic pair production $\gamma \rightarrow e^{+} e^{-}$  is close to unity for  $R=4R{_\odot}$ for a 100 EeV photon in equatorial incidence, whereas for lower energies of about 10 EeV, the conversion probability is close to unity around $R=4R{_\odot}$. Similarly, for polar incidence the same conclusions apply, with slightly higher probability values for a fixed $R$. 

Footprint sizes are found to be enormous. The SPS footprints at a distance of 1AU from the Sun for $R$ varying from $1R{_\odot}$ to $4R{_\odot}$ are as large as $10^{9}$ km and  $10^{4}$ km, respectively. Confrontation of the dipole and DQCS models reveals a difference up to 3 orders of magnitude depending on the $R$ value, with the former producing larger footprints for polar incidence of particles.

\begin{figure}
\centering
\includegraphics[width=0.8\textwidth]{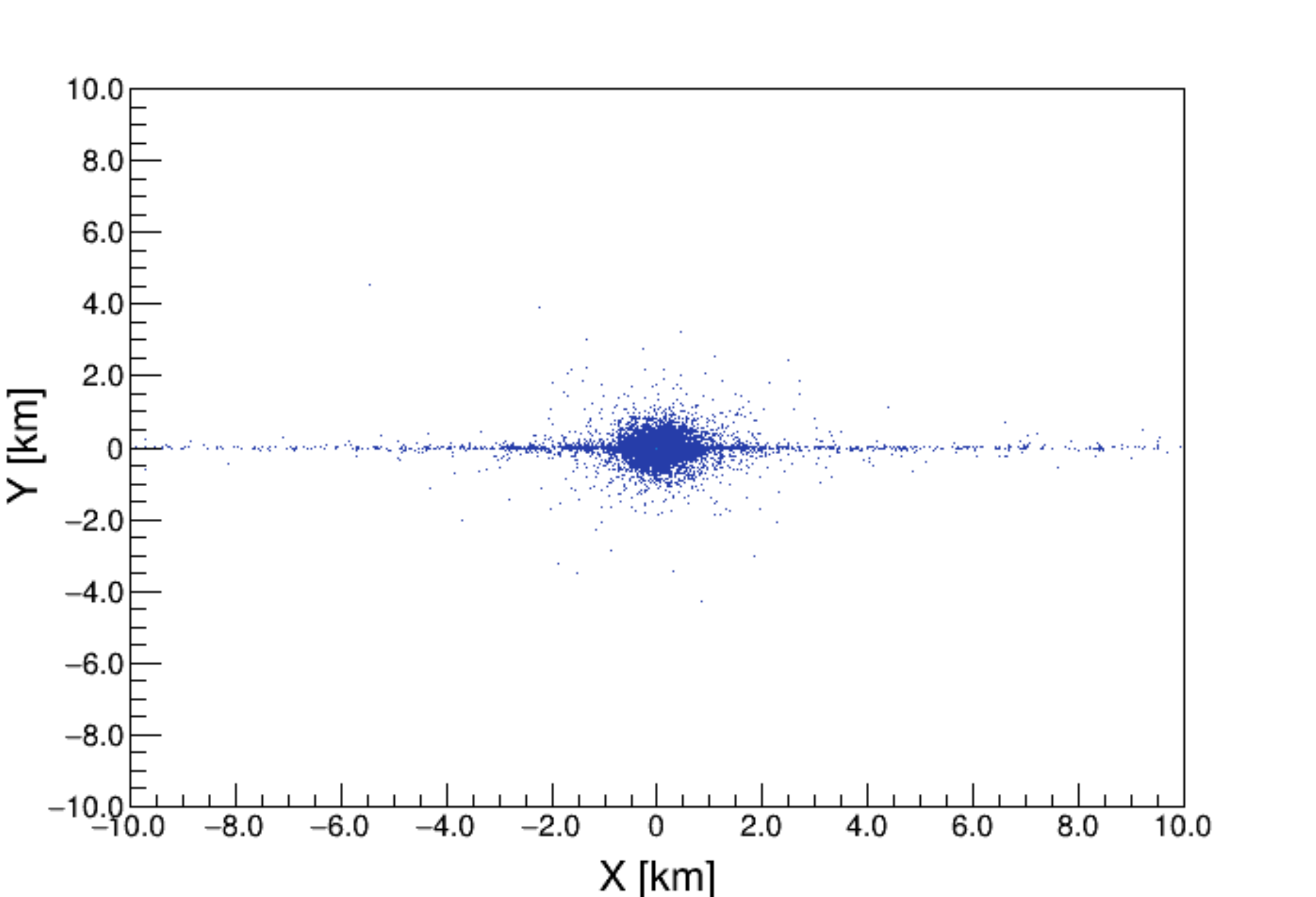}%
\caption{Cosmic Ray Ensemble footprint derived using CORSIKA simulation code for particles that are tracked down through the atmosphere, reacting with air nuclei. Its characteristic topology features a dense, elliptical core accompanied of an extended linear distribution, resulting in a \textit{edge-on spiral} galaxy-like shape.}
\label{fig:footprint}
\end{figure}

\begin{figure*}[!bpht] 
\begin{center}$
\begin{array}{cc}
\includegraphics[width=0.5\textwidth]{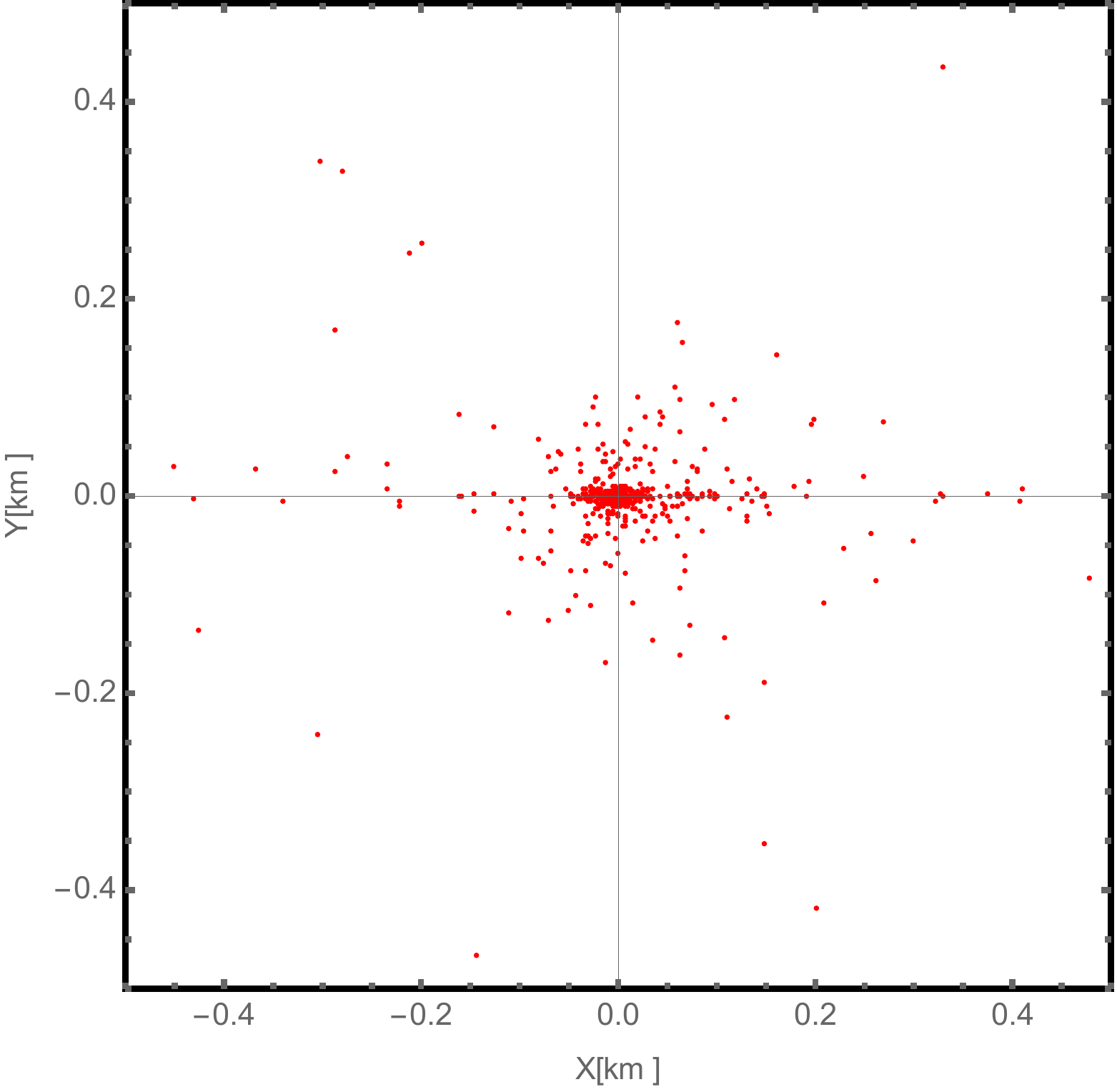}\\
\end{array}$
\end{center} 
\textcolor{blue}{
\caption{Simulation of the recovered footprint distributions produced by a $10^{19}$ eV primary photon within a 1 km$^{2}$ area.
The extension of the events distribution has a length of 0.928 km.}
\label{fig:galaxy_plot}}
\end{figure*}

Figure~\ref{fig:footprint} shows the CRE footprint derived from the simulations from the PRESHOWER and CORSIKA codes for the production and development of the particle distribution going through the atmosphere and finally reaching out the ground level. Although elliptical distributions can be produced from inclined showers, the SPS footprint elliptical core can be potentially identified.
Consequently, the peculiar galaxy-like footprint shape in the detection simulations can be clearly appreciated, with the most energetic particles being located in its core.
Since our goal is to recover this incoming footprint with the highest possible resolution utilizing a detector array, we consider ideal conditions in our simulation: each particle that falls inside a square size detector with maximum efficiency which is equal to unity. As expected, larger areas produce the best results. Figure~\ref{fig:galaxy_plot} shows an example of a recovered footprint distribution produced from a UHE photon of $10^{19}$ eV. Each detector has an area of $25$ cm$^{2}$ and is $25$ cm away from the rest inside the $1$ km$^{2}$ array.

\section{Outlook}

In this study we have found that the simulated CRE footprint can be recovered with a large array of detectors under ideal efficiency conditions. A follow up study will consider more realistic conditions like complex detector configurations or the effect of \textit{unthinned showers}. It is therefore clear that large observatories are best suited for detection of the extended CRE footprints, with facilities like the Pierre Auger Observatory~\cite{2015172}, the Telescope Array or the CREDO network of detectors~\cite{CREDO:2020pzy} already in operation. Consider the surface of detectors of he Pierre Auger Observatory of about 3000 km$^{2}$ which typically collects 3000 cosmic-ray events more energetic than 10 EeV per year. This in fact result in about 0.002 events per year from our considered region around the Sun. Additionally, SPS can be produced from boosted emissions due to gamma-ray bursts from point sources producing a flux one order of magnitude higher than the limits put by the Telescope Array and he Pierre Auger Observatory. In this case, the estimated number of events can be as high as 19 per year~\cite{AlmeidaCheminant:2020sfw}. All in all,  although the expected CRE rate is not certain, we rely on the demonstrated feasibility of detection and verification which can open the door to study diverse physical phenomena of the Universe.

\begin{acknowledgments}
We acknowledge supercomputer support from ACC Cyfronet AGH-UST. This work was partly funded
by the International Visegrad Fund Grant No. 21720040 and by the National Science Centre Grants No. 2016/23/B/ST9/01635 and 2020/39/B/ST9/01398. 
D. A-C. acknowledges support from the Bogoliubov-Infeld program for collaboration between JINR and Polish Institutions.
\end{acknowledgments}

\newpage

\section*{Full Authors List: \Coll\ Collaboration}
%
%
\scriptsize
\noindent
Oleksandr Sushchov$^1$, 
Piotr Homola$^1$, 
David E. Alvarez Castillo$^{1,2}$, 
Dmitriy Beznosko$^3$,
Nikolai Budnev$^4$,
Dariusz G{\'o}ra$^1$,
Alok C. Gupta$^5$,
Bohdan Hnatyk$^6$,
Marcin Kasztelan$^7$,
Peter Kovacs$^8$,
Bartosz {\L}ozowski$^9$,
Mikhail~V.~Medvedev$^{10,11}$,
Justyna Miszczyk$^1$,
Alona Mozgova$^6$,
Vahab Nazari$^{2,1}$,
Micha\l{} Nied{\'z}wiecki$^{12}$,
Maciej Pawlik$^{13,14}$,
Mat{\' i}as Rosas$^{15}$,
Krzysztof Rzecki$^{14}$,
Katarzyna Smelcerz$^{12}$,
Karel Smolek$^{16}$,
Jaros\l{}aw Stasielak$^{1}$,
S\l{}awomir Stuglik$^{1}$,
Manana Svanidze$^{17}$,
Arman Tursunov$^{18}$,
Yuri Verbetsky$^{17}$,
Tadeusz Wibig$^{19}$,
Jilberto Zamora-Saa$^{20}$,
Bo\.zena  Poncyljusz$^{21}$,
Justyna M\k{e}drala$^{22}$,
Gabriela Opi{\l}a$^{22}$,
{\L}ukasz Bibrzyck$^{23}$,
Marcin Piekarczyk$^{23}$.\\

\noindent
$^1$Institute of Nuclear Physics Polish Academy of Sciences, Radzikowskiego 152, 31-342 Krak{\'o}w, Poland.\\
$^2$Joint Institute for Nuclear Research, Dubna, 141980 Russia.\\ 
$^3$Clayton State University, Morrow, Georgia, USA.\\
$^4$Irkutsk State University, Russia.\\
$^5$Aryabhatta Research Institue of Observational Sciences (ARIES), Manora Peak, Nainital 263001, India.\\
$^6$Astronomical Observatory of Taras Shevchenko National University of Kyiv, 04053 Kyiv, Ukraine.\\
$^7$National Centre for Nuclear Research, Andrzeja Soltana 7, 05-400 Otwock-{\'S}wierk, Poland.\\
$^8$Institute for Particle and Nuclear Physics, Wigner Research Centre for Physics, 1121 Budapest, Konkoly-Thege Mikl{\'o}s {\'u}t 29-33, Hungary.\\
$^9$Faculty of Natural Sciences, University of Silesia in Katowice, Bankowa 9, 40-007 Katowice, Poland.\\
$^{10}$Department of Physics and Astronomy, University of Kansas, Lawrence, KS 66045, USA.\\
$^{11}$Laboratory for Nuclear Science, Massachusetts Institute of Technology, Cambridge, MA 02139, USA.\\
$^{12}$Department of Computer Science, Faculty of Computer Science and Telecommunications, Cracow University of Technology, Warszawska 24, 31-155  Krak{\'o}w, Poland.\\
$^{13}$ACC Cyfronet AGH-UST, 30-950 Krak{\'o}w, Poland.\\
$^{14}$AGH University of Science and Technology, Mickiewicz Ave., 30-059 Krak{\'o}w, Poland.\\
$^{15}$Liceo 6 Francisco Bauz{\' a}, Montevideo, Uruguay.\\
$^{16}$Institute of Experimental and Applied Physics, Czech Technical University in Prague.\\
$^{17}$E. Andronikashvili Institute of Physics under Tbilisi State University, Georgia.\\
$^{18}$Research Centre for Theoretical Physics and Astrophysics, Institute of Physics, Silesian University in Opava, \\ Bezru{\v c}ovo n{\'a}m. 13, CZ-74601 Opava, Czech Republic.\\
$^{19}$University of {\L}{\'o}d{\'z}, Faculty of Physics and Applied Informatics, 90-236 {\L}{\'o}d{\'z}, Pomorska 149/153, Poland.\\
$^{20}$ Universidad Andres Bello, Departamento de Ciencias Fisicas, Facultad de Ciencias Exactas, Avenida Republica 498, Santiago, Chile.\\
$^{21}$ Faculty of Physics, University of Warsaw, 02-093 Warsaw, Poland.\\
$^{22}$ Faculty of Physics and Applied Computer Science, AGH University of Science and Technology, 30-059 Cracow, Poland.\\
$^{23}$ Pedagogical University of Krakow, Institute of Computer Science, ul. Podchor\k{a}\.zych, 30-084 Krak{\'o}w, Poland.

%
%
%

\end{document}